\documentclass[prl,twocolumn,showpacs,superscriptaddress]{revtex4}
\usepackage{graphicx,psfrag,amsmath,amssymb,amsfonts,bbm,latexsym,color}

\begin{document}

\title{Reaching Fermi degeneracy in two-species optical dipole traps}

\author{Roberto Onofrio}
\affiliation{Dipartimento di Fisica ``G. Galilei,''
 Universit\`a di Padova, Via Marzolo 8, Padova 35131, Italy}
\affiliation{Istituto Nazionale per la Fisica della Materia, 
Unit\`a di Roma 1 and Center for Statistical Mechanics and Complexity}
\affiliation{Los Alamos National Laboratory, Los Alamos, New Mexico 87545}
\author{Carlo Presilla}
\affiliation{Dipartimento di Fisica, Universit\`a di Roma ``La Sapienza,''
Piazzale A. Moro 2, Roma 00185, Italy}
\affiliation{Istituto Nazionale per la Fisica della Materia, 
Unit\`a di Roma 1 and Center for Statistical Mechanics and Complexity}
\affiliation{Istituto Nazionale di Fisica Nucleare, Sezione di Roma 1}
\date{\today}

\begin{abstract}
We propose the use of a combined optical dipole trap to achieve Fermi 
degeneracy by sympathetic cooling with a different bosonic species. 
Two far-detuned pairs of laser beams focused on the atomic clouds are 
used to confine the two atomic species with different trapping strengths. 
We show that a deep Fermi degeneracy regime can be potentially achieved 
earlier than Bose-Einstein condensation, as discussed in the favorable 
situation of a $^6$Li--$^{23}$Na mixture. 
This opens up the possibility of experimentally investigating a mixture
of superfluid Fermi and normal Bose gases.
\end{abstract}

\pacs{05.30.Fk, 32.80.Pj, 67.60.-g, 67.57.-z}
\maketitle

The study of ultracold atomic gases has witnessed in the last decade an 
impressive development mainly originated from the successful application 
of new experimental techniques such as laser \cite{Chu,Cohen,Phillips} 
and evaporative cooling \cite{Ketterle}. 
The interplay of these two cooling mechanisms has led to the degeneracy 
regime for Bose gases \cite{Wieman}, and to the related rich 
phenomenology recently explored \cite{Dalibard,Pethick}. 
However, these cooling techniques cannot be straightforwardly extended to 
reach degeneracy in Fermi dilute gases, 
and the status of their exploration is by far less advanced than 
for Bose systems. 
One limitation of current efforts to reach Fermi degeneracy arises 
from the use of magnetic trapping techniques, 
where spin-polarized Fermi gases are obtained. 
The Pauli principle limits the efficiency of direct evaporative cooling 
among fermions in the same hyperfine state, and also inhibits scattering 
among fermions in different hyperfine states (Pauli blocking) \cite{DeMarco}.  
An alternative technique - sympathetic cooling of fermions through coupling 
to an ultracold bosonic reservoir \cite{Truscott,Schreck,Hadzibabic} - is 
limited by the decreased efficiency of the elastic scattering 
between fermions and bosons expected when the latter enter a superfluid 
regime \cite{Timmermans,Chikkatur}, and ultimately by Pauli blocking.
Moreover, the use of magnetic trapping strongly limits the kind of 
hyperfine states suitable for obtaining Fermi degeneracy, 
and interferes with 
the use of tunable homogeneous magnetic fields aimed at modulating the 
elastic scattering length via Feshbach resonances \cite{Feshbach}. 
Due to the possibility to unveil new phenomena, 
like a BCS transition to a 
fermionic superfluid state expected at very low temperatures 
\cite{Stoof,Timmermans2001,Holland,Chiofalo}, 
it is crucial to further explore 
efficient cooling techniques for fermions.  

More flexible trapping tools are offered by optical dipole traps 
\cite{Ashkin}. These allow both trapping of different hyperfine states 
and the application of an arbitrary magnetic field. 
While preliminary studies on optical trapping of degenerate Bose gases 
were performed by transferring the condensate from a magnetic trap 
\cite{Stamper}, Bose-Einstein condensation (BEC) \cite{Chapman} 
and Fermi degeneracy \cite{Granade} have been achieved 
directly in all-optical traps, and Feshbach resonances for 
fermions have been studied in an optical dipole 
trap loaded from a magnetic trap \cite{Loftus}. 
It is the purpose of this Letter to discuss a further advantage of using 
an optical dipole trap to achieve Fermi degeneracy, 
based on the possibility 
of trapping different species with selective confinement strengths. 
We discuss the case of an optical dipole trap obtained 
with the combination 
of two laser beams resulting in different trapping potentials for the 
two species of a mixture of Bose and Fermi gases. 
The trapping potentials can be engineered to make the Fermi gas more 
strongly confined than the Bose gas and, therefore, 
the Fermi temperature higher than the BEC critical temperature.
In this way, Fermi degeneracy may be reached well before BEC occurs, 
leading to a promising avenue to look for superfluidity in 
dilute fermions in the presence of a normal Bose gas. 
This would provide an unprecedented situation as compared to the 
already available $^3$He--$^4$He mixtures.

A limitation of the efficiency of sympathetic cooling arises from the 
suppression of elastic scattering of fermions by the bosonic 
reservoir below 
the Bose-Einstein transition temperature \cite{Timmermans}. 
This can be circumvented if the Fermi temperature $T_\mathrm{F}$ is made 
significantly higher than the BEC transition temperature 
$T_\mathrm{c}$ of the bosons. 
Both temperatures scale in the same manner for Fermi and Bose 
atomic gases confined in the same harmonic potential, 
just differing by a numerical factor, 
\begin{eqnarray}
T_\mathrm{F} &=& 1.82~\hbar \omega_\mathrm{f}
~N_\mathrm{f}^\frac{1}{3} k_\mathrm{B}^{-1}
\label{Tf}\\ 
T_\mathrm{c} &=& 0.94~\hbar \omega_\mathrm{b} 
~N_\mathrm{b}^\frac{1}{3} k_\mathrm{B}^{-1},
\label{Tc}
\end{eqnarray} 
with 
$\omega_\mathrm{f}=
(\omega_{\mathrm{f}x} \omega_{\mathrm{f}y} \omega_{\mathrm{f}z})^{1/3}$
($\omega_\mathrm{b}=
(\omega_{\mathrm{b}x} \omega_{\mathrm{b}y} \omega_{\mathrm{b}z})^{1/3}$)
being the geometrical average of the angular trapping 
frequencies in the three directions for fermions (bosons), 
$N_\mathrm{f}$ and $N_\mathrm{b}$ the number of atoms of the Fermi 
and Bose gases, and $\hbar$ and $k_\mathrm{B}$ the Planck and Boltzmann 
constants, respectively. 
In a magnetic trap, since the magnetic moments of the alkali-metal atoms are very 
similar, the only difference in the trapping frequencies of different 
atomic species is due to the difference in their mass. 
In an optical dipole trap instead, the trapping frequencies for 
different atomic species can be made different by orders of magnitude by 
also properly choosing the laser beams.

\begin{figure}[t]
\begin{center}
\includegraphics[width=0.90\columnwidth]{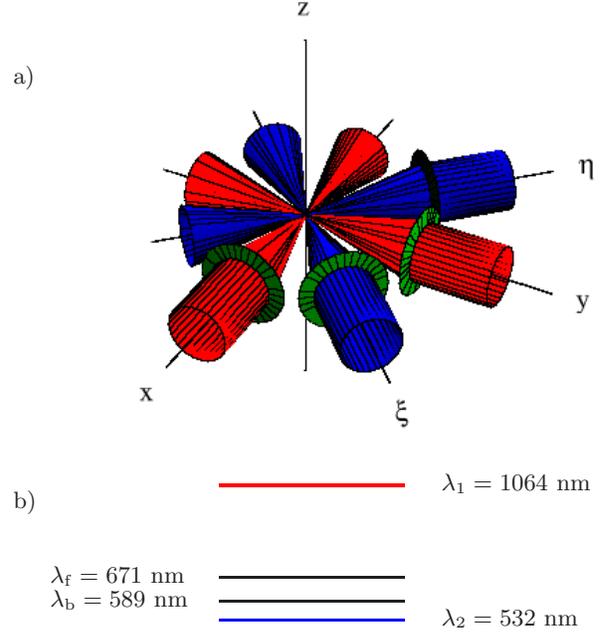}
\flushleft{\vspace{-55mm}a)}
\flushleft{\vspace{50mm}b)}
\vspace{-10mm}
\begin{center}
{
\unitlength=0.2pt
\begin{picture}(450.00,300.00)(0.00,0.00)
\put(420.00,5.00){\makebox(0.00,0.00)[cl]{$\lambda_2=532~\mathrm{nm}$}}
\put(-320.00,40.00){\makebox(0.00,0.00)[cl]{$\lambda_\mathrm{b}=589~\mathrm{nm}$}}
\put(-320.00,85.00){\makebox(0.00,0.00)[cl]{$\lambda_\mathrm{f}=671~\mathrm{nm}$}}
\put(420.00,260.00){\makebox(0.00,0.00)[cl]{$\lambda_1=1064~\mathrm{nm}$}}
\put(0.00,255.00){\color[rgb]{1,0,0}{\rule{350.00\unitlength}{4.00\unitlength}}}
\put(0.00,80.00){\rule{350.00\unitlength}{4.00\unitlength}}
\put(0.00,35.00){\rule{350.00\unitlength}{4.00\unitlength}}
\put(0.00,0.00){\color[rgb]{0,0,1}{\rule{350.00\unitlength}{4.00\unitlength}}}
\end{picture}}
\end{center}
\caption{Combined optical dipole trap for two-species mixtures. 
a) Schematic of the geometry of the two pairs of red- and blue-detuned 
laser beams propagating along the orthogonal axes $x$-$y$ and 
$\xi$-$\eta$, depicted for the case $\theta=\pi/4$. 
b) Relevant atomic wavelengths for the $D_2$ lines in the case of 
a ${}^6$Li-${}^{23}$Na mixture and laser wavelengths for the optical 
trapping.}
\label{fig:set-up}
\end{center}
\end{figure}

The trap we discuss in the following is sketched in Fig. 
\ref{fig:set-up}. 
A pair of laser beams is focused on the center of the trapping 
potential in a crossed-beam geometry \cite{Adams} and provides, 
if red-detuned with respect to the atomic transitions, 
an effective attractive potential for both the atomic species. 
Another pair of blue-detuned laser beams is added, for instance along the 
same directions or rotated by $45^\circ$ in the same plane formed by the 
first two beams, and focused on the same point. 
This second pair is used to weaken the attractive potential created by 
the first one in a selective way. 
The basic idea is that,
due to the different detunings experienced by the two species, 
the combination of the effective potentials resulting from the laser 
beams will give rise to a weaker confinement for one species with 
respect to the other one. 
If the confinement is stronger for the fermionic species, the degeneracy 
condition for it will be met earlier than for the more weakly 
confined bosonic species. 
The situation seems particularly favorable in the case of a 
$^6$Li--$^{23}$Na mixture (see Fig. \ref{fig:set-up}), 
which has been recently brought to a 
degenerate regime in a magnetic trap \cite{Hadzibabic}.  
The two laser wavelengths are chosen at $\lambda_1=1064$ nm 
and $\lambda_2=532$ nm, for instance by using a Nd:YAG laser 
and a frequency-doubling crystal, respectively. 
The relevant atomic transition for sodium is at 
$\lambda_\mathrm{b}=589$ nm 
while for lithium is at $\lambda_\mathrm{f}=671$ nm. 
With respect to the sodium atoms, the lithium atoms are closer to the 
(attractive) red-detuned laser at 1064 nm and farther from the 
(repulsive) blue-detuned laser at 532 nm.     

With the coordinate system chosen in Fig. \ref{fig:set-up}, the effective 
potential energy felt by an atom of species $\alpha$ 
($\alpha=\mathrm{b}$ for $^{23}$Na and $\alpha=\mathrm{f}$ for $^{6}$Li) 
and due to the laser beams $i$ ($i=1,2$) is
\cite{Ashkin}
\begin{eqnarray}
\label{potential}
U_{i}^{\alpha}(x,y,z)
&=&
-\frac{\hbar \Gamma_{\alpha}^2}{8 I_{\alpha}^\mathrm{sat}}
\left( \frac{1}{\Omega_\alpha - \Omega_i} +
       \frac{1}{\Omega_\alpha + \Omega_i} \right) 
\nonumber\\&&\times 
I_i(x,y,z),
\end{eqnarray}
where $\Gamma_{\alpha}$ is the atomic transition linewidth, 
$\Omega_\alpha=2 \pi c/\lambda_{\alpha}$, $\Omega_i=2 \pi c/\lambda_i$, 
$I_i$ is the laser intensity, 
and $I_{\alpha}^\mathrm{sat}$ is the saturation 
intensity for the atomic transition, 
expressed in terms of the former quantities as 
$I_{\alpha}^\mathrm{sat}=\hbar \Omega_\alpha^3 \Gamma_\alpha/12 \pi c^2$.
Each laser intensity $I_i$ is the incoherent sum (obtained by proper 
polarization or a relative detuning of the orthogonal beams) of the 
intensities of the two beams propagating along orthogonal directions 
in the $xy$ plane and focused at $(x,y,z)=(0,0,0)$.
According to Fig. \ref{fig:set-up}, we assume that the red-detuned beams
propagate along the axes $x$-$y$ while the blue-detuned ones along 
the axes $\xi$-$\eta$ rotated with respect to $x$-$y$ by an angle 
$\theta$, 
$\xi=x\cos\theta + y\sin\theta$ and $\eta=y\cos\theta - x\sin\theta$ 
with $0\leq \theta \leq \pi/4$.
In both cases, we can write (with $\theta=0$ for the red-detuned beams)
\begin{eqnarray}
\label{intensity} 
I_i(x,y,z) &=& 
\frac{2 P_i}{\pi w_i^2 \left(1+\frac{\xi^2}{R_i^2}\right)}
\exp\left[
-\frac{2(\eta^2+z^2)}{w_i^2 \left(1+\frac{\xi^2}{R_i^2}\right)}\right]
\nonumber \\
&+&
\frac{2 P_i}{\pi w_i^2 \left(1+\frac{\eta^2}{R_i^2}\right)}
\exp\left[
-\frac{2(\xi^2+z^2)}{w_i^2 \left(1+\frac{\eta^2}{R_i^2}\right)}\right],~~~
\end{eqnarray}
where $P_i$ is the beam power, $w_i$ is the $1/e^2$ beam waist radius, 
and $R_i=\pi w_i^2/\lambda_i$ is the Rayleigh range.

The total potential experienced by the fermions (bosons) is 
$U_\mathrm{f}=U_1^\mathrm{f}+U_2^\mathrm{f}$ 
($U_\mathrm{b}=U_1^\mathrm{b}+U_2^\mathrm{b}$). 
For $P_1/P_2$ large enough, both the potentials 
$U_\mathrm{f}$ and $U_\mathrm{b}$ present a minimum at $(x,y,z)=(0,0,0)$.
According to (\ref{Tf}) and (\ref{Tc}), 
the Fermi and the BEC critical temperatures
are determined by the small oscillation frequencies around this minimum.
Neglecting the terms $(\lambda_i/\pi w_i)^2$ with respect to unity, 
we find
\begin{eqnarray}
\omega_{\alpha x} = 
\omega_{\alpha y} = 
\frac{\omega_{\alpha z}}{\sqrt{2}} = 
\sqrt{\frac{\hbar}{2 \pi m_\alpha}
\left(
\frac{k_1^\alpha P_1}{w_1^4}
+
\frac{k_2^\alpha P_2}{w_2^4}
\right)},
\label{omega}
\end{eqnarray}
where $m_\alpha$ is the mass of an atom of the species $\alpha$ and
\begin{eqnarray}
k_{i}^{\alpha} = 
\frac{\Gamma_{\alpha}^2}{I_{\alpha}^\mathrm{sat}}
\left( \frac{1}{\Omega_\alpha - \Omega_i} +
       \frac{1}{\Omega_\alpha + \Omega_i} \right).
\end{eqnarray}
Note that the trapping angular frequencies in (\ref{omega}) do not 
depend on the rotation angle $\theta$ between the blue- and 
red-detuned laser beams, as a consequence of the rotational 
invariance of the potential around the local minimum.

\begin{figure}[t]
\begin{center}
\psfrag{x}[c]{$P_2/P_1$}
\psfrag{y}[c]{$\omega_\mathrm{f}/\omega_\mathrm{b}$}
\includegraphics[width=0.85\columnwidth]{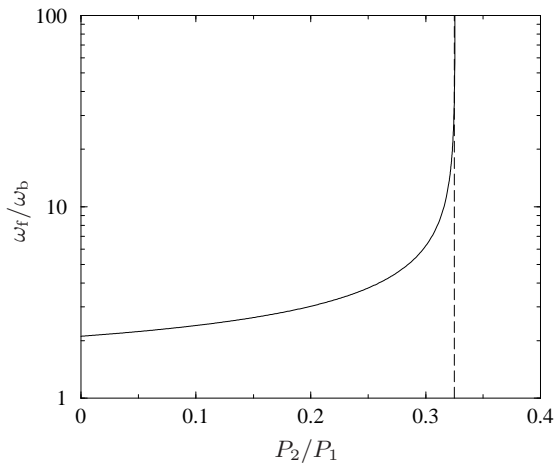}
\caption{
Ratio between the average trapping angular frequencies 
for the fermionic and the bosonic species as a function of 
the beam power ratio between the blue- and red-detuned lasers. 
Note that $\omega_\mathrm{f}/\omega_\mathrm{b}>1$ for $P_2/P_1=0$
as a consequence of the mass ratio between $^6$Li and $^{23}$Na. 
We assume equal waists for the beams, and  
$\Gamma_\mathrm{f}=2 \pi \times 5.9~\mathrm{MHz}$,
$\Gamma_\mathrm{b}=2 \pi \times 9.8~\mathrm{MHz}$,  
$m_\mathrm{f}=9.6 \times 10^{-27}~\mathrm{Kg}$,
$m_\mathrm{b}=36.8 \times 10^{-27}~\mathrm{Kg}$
for the fermionic ($^6$Li)  and bosonic ($^{23}$Na) species, respectively.
The dashed line indicates the analytical prediction (\ref{soglia}) 
for the critical power ratio corresponding to unconfined bosons.}
\label{fig:freq-ratio}
\end{center}
\end{figure}

In Fig. \ref{fig:freq-ratio} we show the ratio between the average 
angular frequencies for the fermionic and the bosonic species,
$\omega_\mathrm{f}=
(\omega_{\mathrm{f}x}\omega_{\mathrm{f}y}\omega_{\mathrm{f}z})^{1/3}$
and 
$\omega_\mathrm{b}=
(\omega_{\mathrm{b}x}\omega_{\mathrm{b}y}\omega_{\mathrm{b}z})^{1/3}$, 
as a function of the beam power ratio between the blue- and 
red-detuned lasers.
The relative confinement becomes progressively stronger for the fermion 
species as the beam power ratio approaches the critical value 
$P_2/P_1 \simeq 0.326$ at which $\omega_\mathrm{b}=0$.
By using (\ref{omega}), we find the following expression for the critical 
power ratio
\begin{equation}
\label{soglia}
\left. \frac{P_2}{P_1} \right|_\mathrm{crit} = 
\frac{\Omega_2^2-\Omega_\mathrm{b}^2}
{\Omega_\mathrm{b}^2-\Omega_1^2}
\left( \frac{w_2}{w_1} \right)^4.
\end{equation}
The strong dependence of the critical power ratio upon the beam waists 
can be used to reduce the amount of blue-detuned light necessary to 
completely deconfine the boson gas.
In terms of absolute values, the trapping frequency in the $z$-direction 
for the bosons in the presence of the red-detuned laser beam alone 
($P_2/P_1=0$) and a beam waist $w_1=10 \mu$m is 
$\nu_{\mathrm{b}z}=\omega_{\mathrm{b}z}/2\pi \simeq 15.87~
P_1^{1/2}~\mathrm{KHz}$, with $P_1$ in Watts. 
Thus, the optical dipole trap allows for large absolute degeneracy 
temperatures comparable or superior to the largest values obtained 
with ingenious designs of magnetic traps \cite{Schreck1}, 
and also mitigates 
any loss of efficiency in the evaporative cooling attributable to the 
differential gravitational sagging of the two trapped species.
Moreover, the presence of a stiffer confinement for the fermions also 
increases the efficiency of evaporative cooling near Fermi 
degeneracy, because it always maintains a large overlap with the less 
confined bosonic species and the latter, unlike the fermion cloud, 
progressively shrinks in size.  

\begin{figure}[t]
\begin{center}
\psfrag{x}[c]{$P_2/P_1$}
\psfrag{y}[c]{$\Delta U/P_1$ ($\mu$K/W)}
\psfrag{fermions}[c]{$^6$Li}
\psfrag{bosons}[l]{$^{23}$Na}
\includegraphics[width=0.85\columnwidth]{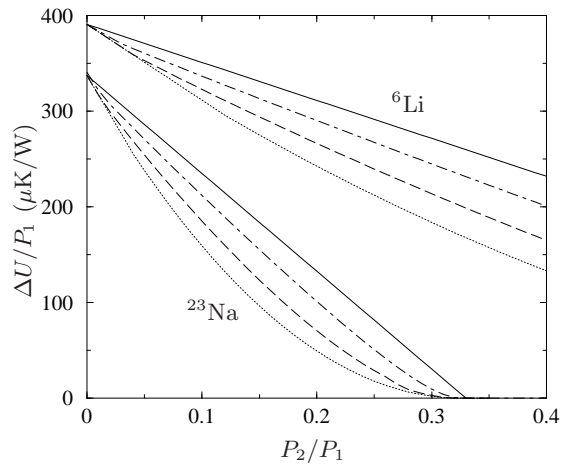}
\caption{Confining energy $\Delta U$ per unit of infrared laser power 
for the fermionic and bosonic species as a function of the beam power 
ratio between the blue- and red-detuned lasers. 
For each species we show from top to bottom the curves obtained 
with the blue-detuned beams rotated with respect to the red-ones 
by $\theta=0$, $\pi/16$, $\pi/8$, and $\pi/4$, respectively.}
\label{fig:conf-energy}
\end{center}
\end{figure}

To check for the confinement feature corresponding to various beam power 
ratios, we have also studied the minimal potential energy depth 
(confining energy) $\Delta U$ of the exact potentials $U_\mathrm{f}$ 
and $U_\mathrm{b}$. 
The behavior of $\Delta U$ for the fermionic and bosonic species 
is shown in Fig. \ref{fig:conf-energy} for four possible angles $\theta$ 
between the red-detuned and the blue-detuned beam pairs.
We see that bosons are always less confined than fermions, and that the 
confinement is weaker for both species when $\theta \neq 0$. 
Thus evaporative cooling of $^{23}$Na does not affect significantly the 
confinement of $^6$Li. 

Based on Fig. \ref{fig:conf-energy}, 
we can imagine the following evaporative cooling dynamics.
First, the red-detuned power 
$P_1$ is decreased as typically done in an optical 
dipole trap \cite{Chapman,Granade,Adams}. 
When the temperature is approaching $T_\mathrm{F}$, 
the blue-detuned laser is turned on 
with the ratio $P_2/P_1$ kept constant 
during the following stage of evaporation.
There is a trade-off in choosing the final ratio $P_2/P_1$ 
since the shallower confinement of the bosons also results 
in a smaller peak density of this species, hence a smaller 
elastic scattering rate, affecting both evaporative cooling and 
the subsequent sympathetic cooling of the fermions. 
With a power ratio $P_2/P_1 \simeq 0.31$, 
equal waists of the laser beams, 
and $N_\mathrm{f}=N_\mathrm{b}$, 
according to (\ref{Tf}) and (\ref{Tc}),
a minimum temperature 
$T \simeq 5 \times 10^{-2} T_\mathrm{F}$ can be achieved before 
the bosons condense. 
Deviations from the ideal Fermi-Bose mixture 
are negligible at temperatures above $T_\mathrm{c}$, due to the weak 
interatomic interactions \cite{Amoruso}. 
Thus one can assume ideal density 
profiles for fermions and bosons and
estimate the elastic collision rate at the final stage of evaporation. 
For sodium atoms we have 
$\Gamma_\mathrm{el} \simeq 30$ Hz for $P_1=10$ mW, 
$w_1=10$ $\mu$m, 
and $N_\mathrm{b}=10^6$ \cite{Note}. 
This rate should be compared to the residual Rayleigh scattering 
from the blue-detuned beams which limit the atom lifetimes, 
provided that other heating sources 
such as laser intensity, frequency and beam-pointing stabilities 
are properly optimized. 
An estimate of the Rayleigh scattering rates for ${}^6$Li and 
${}^{23}$Na gives 
$\gamma_1^\mathrm{Li}=5.7 \times 10^{-1}$ Hz, 
$\gamma_1^\mathrm{Na}=4.5 \times 10^{-1}$ Hz, 
$\gamma_2^\mathrm{Li}=7.7$ Hz, and
$\gamma_2^\mathrm{Na}=38$ Hz 
due to the red-detuned and blue-detuned beams, respectively, 
all referred to a laser power of 1 W and 
evaluated at the peak laser intensity. 
The biggest contribution comes from the blue-detuned beam acting 
on sodium atoms, due to the proximity of the corresponding frequencies. 
However, this is of less concern when one considers that the 
blue-detuned beams are only turned on in the latest stage of the 
evaporation, when approaching the Fermi degeneracy regime.
Moreover, their powers are progressively increased to less than 1/3 
of the simultaneous value of the power of the red-detuned beams 
(with $w_2<w_1$ this ratio can be made even smaller according to 
Eq. (\ref{soglia})) while the latter undergo a continuous decrease. 
For a power $P_2=3.1$ mW, in the same conditions 
as above we obtain a Rayleigh scattering rate 
$\gamma_2^\mathrm{Na}=1.2 \times 10^{-1} ~\mathrm{Hz} 
\simeq 1/250~\Gamma_\mathrm{el}$. 
Thus atom lifetimes in excess of about 10 s should be achievable.
These are long enough to perform experiments requiring mechanical 
stirring of the fermion cloud.
The above analysis can be repeated by considering a CO${}_2$ laser as 
red-detuned source.
In this case, we should expect advantages in terms of larger and more 
stable available power, superior beam-pointing stability, 
and smaller residual Rayleigh scattering \cite{Ohara}. 
We plan to discuss this point in detail elsewhere. 

In conclusion, we have outlined  a novel strategy to reach a deep Fermi 
degenerate regime based upon a proper engineering of a two-species 
optical dipole trap. The case of a $^6$Li--$^{23}$Na mixture has been 
discussed in detail also due to the very favorable properties of 
this mixture recently reported in \cite{Hadzibabic}. 
The strategy can be also applied to other mixtures such as 
$^{40}$K--$^{87}$Rb \cite{Goldwin} 
or $^6$Li--$^{87}$Rb for which $\lambda_\mathrm{b} > \lambda_\mathrm{f}$, 
by choosing a blue-detuned beam wavelength such that 
$\lambda_\mathrm{f} < \lambda_2 < \lambda_\mathrm{b}$.
Our proposal could pave the road to the experimental study of a 
novel phase consisting of a superfluid Fermi gas and a normal Bose gas, 
a situation precluded in the $^3$He--$^4$He Fermi-Bose mixtures. 
Besides providing access to a new system interesting in itself, 
this could considerably simplify signatures for fermion superfluidity 
based on the direct imaging of the density profile of the trapped 
fermions \cite{Chiofalo}.

\begin{acknowledgments}
We thank E. Timmermans, D. J. Vieira, L. Viola, and X. Zhao 
for a critical reading of the manuscript. 
This work was supported in part by 
Cofinanziamento MIUR protocollo MM02263577\_001. 
One of us (CP) also acknowledges hospitality from the Theoretical 
Division at the Los Alamos National Laboratory.
\end{acknowledgments}


\begin{references}
\bibitem{Note} 
The fermion-fermion elastic scattering rate is comparable to the
boson-boson one since the suppression factor due to Pauli blocking is 
compensated by the higher density due to the larger trapping frequency. 
For the interspecies elastic collision rate no data are available,
however the accomplishments reported in \cite{Hadzibabic} suggest 
that this process should not be the leading limitation for sympathetic 
cooling.
\end{references}
\end{document}